\documentclass[reprint,aps,pra,superscriptaddress,longbibliography]{revtex4-2}

\usepackage{graphicx}
\usepackage{hyperref}
\usepackage{siunitx}

\usepackage{amsmath,mathtools}
\usepackage{amssymb}

\newcommand{\bra}[1]{\langle #1 \vert}
\newcommand{\ket}[1]{\vert #1 \rangle}
\newcommand{\braket}[2]{\langle #1 | #2 \rangle}

\usepackage[dvipsnames]{xcolor}

\begin{document}

\title{State independent QKD}
\author{Robert Kindler}
\email{robert.kindler@oeaw.ac.at}
\affiliation{Institute for Quantum Optics and Quantum Information (IQOQI), Austrian Academy of Sciences, Boltzmanngasse 3, 1090 Vienna, Austria}
\author{Johannes Handsteiner}
\affiliation{Institute for Quantum Optics and Quantum Information (IQOQI), Austrian Academy of Sciences, Boltzmanngasse 3, 1090 Vienna, Austria}
\author{Jaroslav Kysela}
\affiliation{Institute for Quantum Optics and Quantum Information (IQOQI), Austrian Academy of Sciences, Boltzmanngasse 3, 1090 Vienna, Austria}
\affiliation{Vienna Center for Quantum Science and Technology (VCQ), Faculty of Physics, Boltzmanngasse 5, University of Vienna, Vienna A-1090, Austria.}
\author{Kuntuo Zhu}
\affiliation{Institute for Quantum Optics and Quantum Information (IQOQI), Austrian Academy of Sciences, Boltzmanngasse 3, 1090 Vienna, Austria}
\affiliation{State Key Laboratory of Low-Dimensional Quantum Physics and Department of Physics, Tsinghua University, Beijing 100084, China}
\author{Bo Liu}
\affiliation{Institute for Quantum Optics and Quantum Information (IQOQI), Austrian Academy of Sciences, Boltzmanngasse 3, 1090 Vienna, Austria}
\affiliation{College of Advanced Interdisciplinary Studies, National University of Defense Technology, Changsha 410073, China}
\author{Anton Zeilinger}
\affiliation{Institute for Quantum Optics and Quantum Information (IQOQI), Austrian Academy of Sciences, Boltzmanngasse 3, 1090 Vienna, Austria}


\date{\today} 

\begin{abstract}

We present an adaptive procedure for aligning quantum non-locality experiments without any knowledge of the two-qudit state shared by the participating parties. The quantum state produced by the source, its unitary evolution as well as the actual measurement bases remain unknown to both parties at all times. The entanglement of the quantum state helps establish desired correlations between individual measurement bases of the two distant parties. We implement the procedure in a fiber-based quantum key distribution (QKD) setup with polarization-entangled photons, where we do not rely on any additional alignment tools such as lasers or polarizers. In a QKD scenario the procedure can be done without any additional measurements as those that are performed regardless.
\end{abstract}

\maketitle


\section{Introduction}

\begin{figure*}[!htp]
\begin{center}
\includegraphics[width=0.9\textwidth,clip]{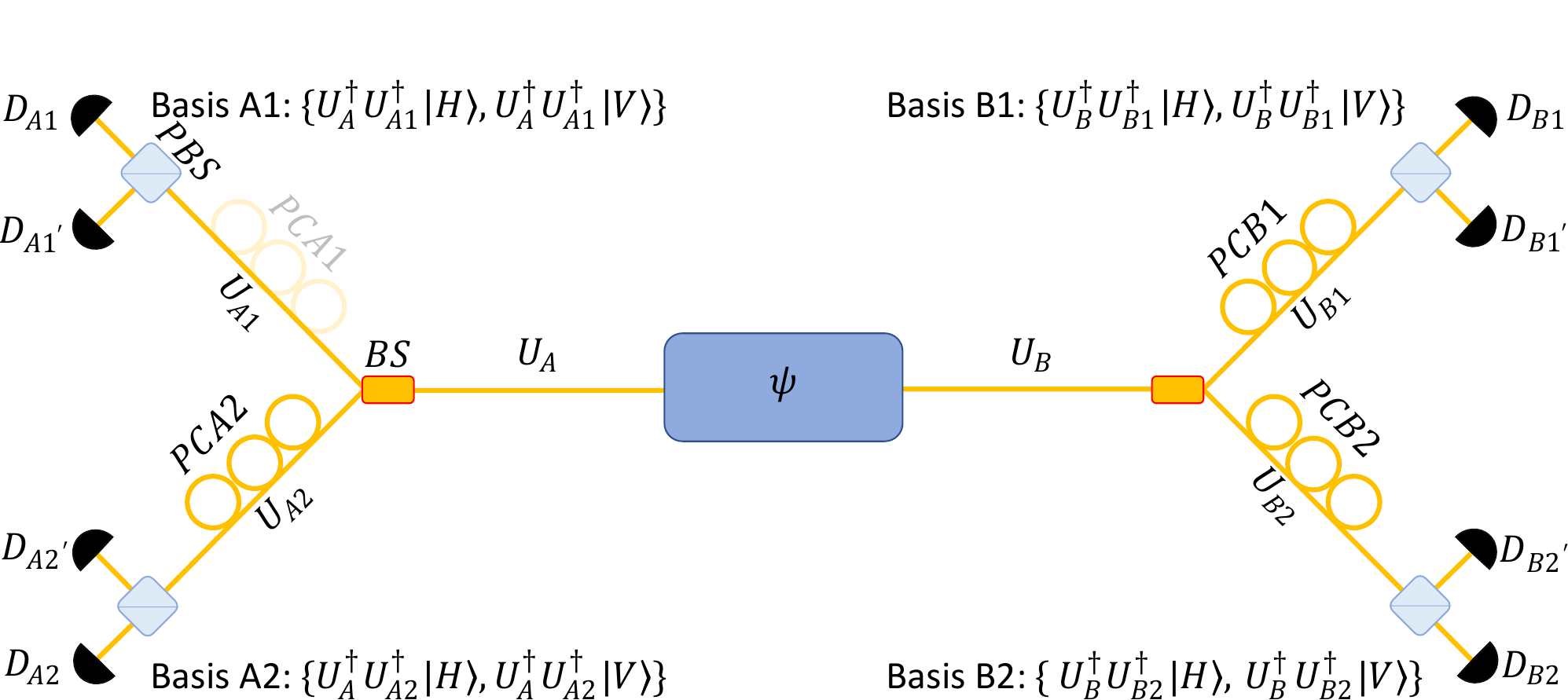}
\end{center}
\caption{The source is of Sagnac type and sends entangled photons to the receivers Alice and Bob. An in-fiber 50/50 beam splitter (BS) randomly routes each photon to one of two measurement bases (A1, A2 for Alice and B1, B2 for Bob), each consisting of a polarizing beam splitter (PBS) followed by two single photon detectors ($D_{A1,2}$, $D_{A1',2'}$, $D_{B1,2}$, $D_{B1',2'}$). The in-fiber polarization controllers~(PCA2, PCB1 and PCB2) were used to align the measurement bases relative to each other in order to fulfill all required conditions on the correlations.}
\label{fig1}
\end{figure*}

Quantum entanglement plays an increasing role not only in the fundamental science but also in the application-driven research. The correlations exhibited by entangled systems lead for instance to violations of Bell inequalities in tests of local realism \cite{bell_speakable_1997} or can be utilized for unconditionally secure communication via quantum key distribution (QKD) protocols \cite{Bennett:1984wv,Scarani_2009}. In these setups, two parties, commonly referred to as Alice and Bob, each hold a part of an entangled state and subject their parts to local quantum measurements in several (usually two) measurement bases. Provided the bases are chosen properly, the measurement results exhibit correlations that cannot be explained by a local realistic description alone. For making use of quantum phenomena, precise alignment of measurement apparatuses of both participating parties is therefore crucial. This is further complicated by disturbing effects of the environment. In polarization-encoded protocols, for example, temperature drifts or mechanical stress in fiber-based QKD systems~\cite{Treiber2009} and rotating reference frames in some free-space settings like satellite-based QKD systems~\cite{Liao2017,Yin2017} require compensation. This adds complexity to the experimental setups and can result in a loss of operating time, when the compensation requires temporary interruption of the key distribution, or photons are lost for key generation due to additional measurements. 

In this paper, we demonstrate how measurement bases in non-locality experiments can be aligned properly even in the absence of a global reference frame and costly compensation techniques. The key is to utilize the very entanglement that is subsequently used to violate the Bell inequalities or to extract a secret key in QKD protocols. In a sense, it is the entanglement itself that serves to establish a common reference frame for both parties. Both Alice and Bob measure in two bases each, which are completely unknown to them and can be different when running the experiment again at a later time. Our alignment procedure only makes sure that Alice's first basis is mutually unbiased to her second basis, while being simultaneously perfectly correlated to Bob's first basis and completely uncorrelated to his second basis (and analogously for Bob). We neither know, nor do we need to know, what these bases actually are and no further information about the shared entangled state is required. 
We demonstrate our alignment procedure for the specific case of fiber-based polarization-encoded two-party QKD BBM92~\cite{BBM92} protocol. Nevertheless, the procedure can be modified also for other degrees of freedom, for higher-dimensional qudits, and is not restricted to QKD settings.

Note that a polarization compensation scheme for the BB84-protocol was demonstrated by Ding et al.~\cite{Ding2017}, and for entanglement-based QKD by Shi et al.~\cite{shi2021fibre}. The latter required a known $\ket{\psi^-}$ state and unbiased measurement bases on each side. In other experiments of this kind, the polarization of the photons needed to be corrected by first measuring the polarization explicitly at the end of the glass fiber and then adjusting it accordingly by using a polarization controller. Sometimes this included the use of reference lasers, sometimes a part of the signal was channeled off. In all of these experiments, the polarization was measured and set explicitly at one point. Additionally, the entangled state of photons was well known and set explicitly by the experimenter~\cite{Weihs1998,Treiber2009,joshi,Wengerowsky}. The main advantage of our approach is the fact that we can disregard all of these methods and tools. We can align our setup only by measuring single counts and coincidence counts on Alice's and Bob's side in two unknown (arbitrary and not characterized) bases. A small fraction of counts is communicated publicly. This fraction can be changed dynamically to minimize the impact on the secure key rate. For the polarization alignment and stabilization we use fully automated polarization controllers, allowing us to perform quantum key distribution for in principle unlimited time, in an plug-and-play scheme without any further alignment and external control. 

\section{Methods}
\label{sec:methods}

The receivers in entanglement-based QKD systems need to perform measurements on pairs of qubits in certain measurement bases. For protocols like BB84~\cite{Bennett:1984wv} and BBM92~\cite{BBM92} it is necessary for each party to measure in two mutually unbiased bases. Alice and Bob both use four detectors each (two for each basis) and write down time-stamps for each detection event. This data is processed in real time and whenever one detector of Alice and one of Bob click at the same time (within a certain short time window), the two events are regarded as a coincidence count and assumed to correspond to one detected photon pair. The alignment procedure presented below is based on monitoring these coincidence count rates for different combinations of measurement bases for Alice and Bob. The rates are used as a feedback signal for adaptive modifications of the measurement bases. A1 and A2 correspond to Alice's first and second basis. Bob's bases are labelled accordingly as B1 and B2. The schematic setup is depicted in Fig. \ref{fig1}. An entangled photon source produces in this case polarization entangled photon pairs and sends them to the receivers Alice and Bob.


Both transmission channels are subjected to local random unitary transformations $U_A$ and $U_B$ respectively which alter the polarization and are slowly changing over time. These changes are out of Alice's and Bob's control and caused by environmental effects. Additionally, Alice and Bob can freely manipulate the transformations $U_{A1} / U_{A2}$ and $U_{B1} / U_{B2}$ after the $50/50$ beam splitter by fiber paddles, but are unaware about the actual mathematical form of these transformations. 
These channels, given by $U_A / U_B$, $U_{A1} / U_{A2}$ and $U_{B1} / U_{B2}$, can either be seen as part of the source or part of the measurement basis. If they are seen as part of the entangled photon source, this means that they transform the source's original state into another maximally entangled state, which is unknown to Alice and Bob. 
If these transmission channels are seen as part of the detection setup, the transformations simply rotate the measurement bases into something unknown to Alice and Bob. The intermediate case is also possible, where unitaries $U_A / U_B$ are included in the source, whereas unitaries $U_{A1} / U_{A2}$ and $U_{B1} / U_{B2}$ rotate the bases. Unlike in the two previous cases however, one cannot make any specific claims about the form of the bases as well as the form of the entangled state, as both include transformations that remain unknown.

With the alignment steps listed below, Alice and Bob ensure that their bases A1 and B1 (i.e., $\{U_A^\dag U_{A1}^\dag\ket{H}, U_A^\dag U_{A1}^\dag\ket{V}\}$ and $\{U_B^\dag U_{B1}^\dag\ket{H}, U_B^\dag U_{B1}^\dag\ket{V}\}$) as well as A2 and B2 (i.e., $\{U_A^\dag U_{A2}^\dag\ket{H}, U_A^\dag U_{A2}^\dag\ket{V}\}$ and $\{U_B^\dag U_{B2}^\dag\ket{H}, U_B^\dag U_{B2}^\dag\ket{V}\}$) are correlated, while the other two combinations (A1 and B2, A2 and B1) are uncorrelated. Furthermore, the procedure guarantees that A1 is mutually unbiased to A2 as well as B1 is mutually unbiased to B2 (for proof see appendix \ref{appendix}):
\begin{equation}
\begin{aligned}
&|\bra{k}U_{A1}U_{A2}^{\dag}\ket{l} |^2=\frac{1}{2} \quad \text{and} \\ 
&|\bra{k}U_{B1}U_{B2}^{\dag}\ket{l}|^2=\frac{1}{2} \quad \text{for }k,l\in \{H,V\} \\
& 
\end{aligned}
\end{equation}
As a feedback signal for our polarization controllers, we calculate the visibilities for all combinations of measurement bases. For example, the visibility $V_{A1,B1}$ between A1 and B1 is defined as
\begin{widetext}
\begin{equation}
\label{eqvis}
V_{A1,B1} = \frac{CC_{DA1,DB1}+CC_{DA1',DB1'}-CC_{DA1,DB1'}-CC_{DA1',DB1}}{CC_{DA1,DB1}+CC_{DA1',DB1'}+CC_{DA1,DB1'}+CC_{DA1',DB1}},
\end{equation}
\end{widetext}
where $CC_{DA1,DB1}$ denotes the coincidence detection rate between the detector DA1 in the transmitted arm in A1 and the detector DB1 in the transmitted arm in B1. All other visibilities and coincidence rates are denoted likewise. The quantum bit error rate $\text{QBER}_{A1,B1}$ can be computed out of the visibility $V_{A1,B1}$ by using the simple formula $\text{QBER}_{A1,B1}=\frac{1-|V_{A1,B1}|}{2}$.

The easiest and fastest alignment procedure comprises the following steps:
\begin{enumerate}
    \item  Maximize the visibility $V_{A1,B1}$ 
  (for instance with the polarization controller PCB1). This is equivalent to trying to get the $\text{QBER}_{A1,B1}$ to $0$.

    \item Minimize the absolute value of the visibility $|V_{A1,B2}|$ between A1 and B2 to zero, i.e., the $\text{QBER}_{A1,B2}$ between A1 and B2 to 50\% (for instance with the polarization controller PCB2).

    \item Maximize $V_{A2,B2}$, i.e, the quantum bit error rate $\text{QBER}_{A2,B2}$ should be set to $0$  (for instance with the polarization controller PCA2).
\end{enumerate}

\begin{figure}[!htp]
\begin{center}
\includegraphics[width=0.5\textwidth,clip]{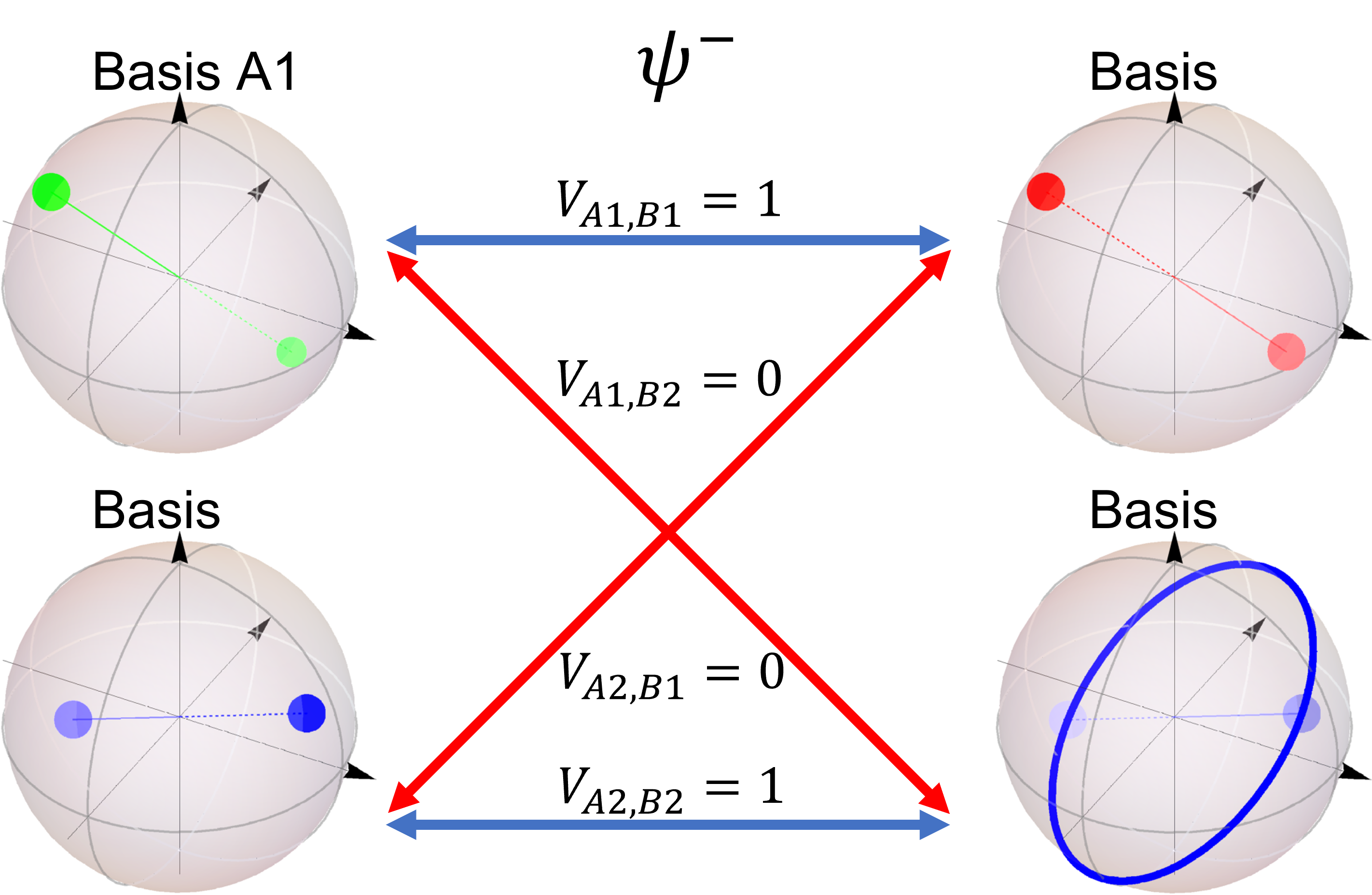}
\end{center}
\caption{The Poincar\'{e} spheres visualize how the measurement bases are fixed by our alignment protocol if Alice and Bob share a $\ket{\psi^-}$ state. Note that the spheres are arbitrarily oriented (but all in the same way). 
A basis corresponds to two antipodal points on the sphere. Basis A1 is never changed and allowed to take any arbitrary position. At the first alignment step B1 is changed to be correlated with A1 and therefore anti-parallel to it. B2 is set to be uncorrelated to A1, which is fulfilled for all bases lying perpendicular to A1 on the blue circle. At last, A2 is set to be correlated and therefore anti-parallel to B2. In case the shared state is not $\ket{\psi^-}$, the procedure works exactly the same but correlating or uncorrelating two bases does not anymore imply they are anti-parallel or perpendicular, respectively.
}
\label{Bild2}
\end{figure}

This procedure is sufficient to align the setup for QKD, see appendix \ref{appendix}, even though Alice and Bob neither know nor do they need to know what exact basis they are measuring in. Alternatively, one can also align this setup the exact same way on anti-correlations by setting $V_{A1,B1}=-1$ and/or $V_{A2,B2}=-1$. An intuitive picture of this can be obtained by assuming the source emits a $\ket{\psi^-}$ state. For this state the visibility only depends on the relative angle between the two measurement bases on the Poincar\'{e} spheres relative to each other and is given by $V^{\psi^{-}}=-\cos(\alpha-\beta)$, whereby we assumed without loss of generality that both bases lay within a plane and are described only by the angles $\alpha$ and $\beta$. This is not restricted to linear polarization but true for any two measurement bases and can intuitively be understood when keeping in mind that the $\ket{\psi^-}$ state is rotation invariant. The final aligned form of the four measurement bases is visualized in Fig. \ref{Bild2}. Note that if and only if the source emits the $\ket{\psi^-}$ state, this also implies that A1$=\pm$B1, A2$=\pm$B2 and A1 is mutually unbiased to B2 as well as A2 is mutually unbiased to B1. Also note that when talking about different bases, we refer to the whole transmission channel from the detectors back to the source (including $U_A$ and $U_B$). Therefore, in this case, the relative position of measurement bases on the Poincar\'{e} sphere is important, while the global orientation of the sphere is irrelevant. 

Without considering the $\ket{\psi^-}$ state, the geometrical picture is not as intuitive anymore as the correlations do not necessarily depend on the the angle between the measurement bases. However, the alignment procedure still works exactly the same way and Alice as well as Bob will end up with two mutually unbiased bases, as long as the input state is maximally entangled. It is noteworthy to point out that in this general case, two uncorrelated bases are not necessarily mutually unbiased and even two identical measurement bases do not need to be correlated. Details and an analytic proof of these statements can be found in appendix \ref{appendix}. 

\section{Results}

We established a proof-of-principle QKD-setup using only the visibility (or QBER) as alignment and stabilization tool. The source is based on a Sagnac interferometer generating polarization-entangled photon pairs with their state of the form:
\begin{equation}
   \ket{\psi}=\frac{1}{\sqrt{2}}\left(\ket{HV}+e^{i\phi}\ket{VH}\right).
\end{equation}
Fiber-based beam splitters were used on each side to choose the measurement bases. We used fully automated, in-fiber piezo-based polarization controllers to change Alice's and Bob's measurement bases. Subsequently, the light was collimated and analyzed by free space polarizing beam splitters~(PBSs). After each output port of the PBS the photons were coupled into multimode fibers and detected by -- in total -- eight single photon detectors. The electrical signals were time-tagged by a time to digital converter and recorded on a PC. A freely choosable fraction of all registered counts was used to calculate visibilities in real time. This information was then used in a feedback loop to adjust the polarization controllers. All polarization controllers were controlled via a homemade LabVIEW program that performed the whole alignment procedure automatically. The setup is schematically displayed in Fig. \ref{fig1}.

\begin{figure}[!htp]
\begin{center}
\includegraphics[width=0.5\textwidth,clip]{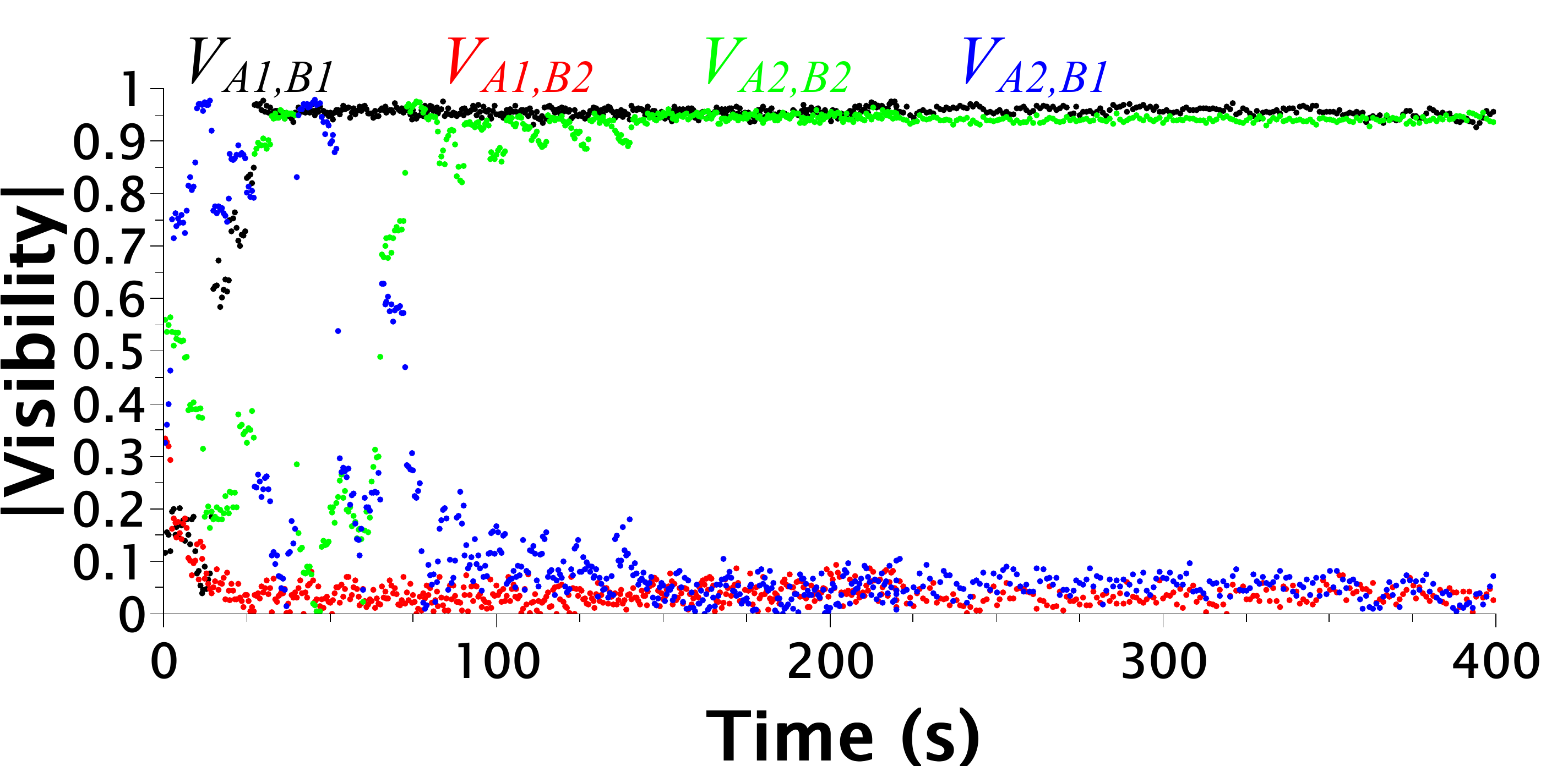}
\end{center}
\caption{Automated alignment procedure of the setup with three polarization controllers, each assigned to align one visibility curve: black: $|V_{A1,B1}|$ (set by PCB1), red: $|V_{A1,B2}|$ (set by PCB2), green: $|V_{A2,B2}|$ (set by PCA2), blue: $|V_{A2,B1}|$. Here we plot the absolute value of each visibility. All polarization controllers were active at the same time and the alignment was stable after around 180 seconds.
}
\label{auto2}
\end{figure}

Fig. \ref{auto2} shows the alignment procedure as described in section \ref{sec:methods}. A maximally polarization-entangled state is distributed between Alice and Bob. The whole procedure was automated by assigning each of the three alignment steps mentioned above to motorized polarization controllers. All three steps were not done in chronological order, but performed simultaneously \footnote{This only works because alignment of $V_{A1,B1}$ (i.e. step 1) cannot be disturbed by PCB2 and PCA2, which are used to perform step 2 and 3. Likewise, $V_{A1,B2}$ (i.e. step 2) cannot be disturbed by PCA2, which is used for step 3.}.

After all alignment conditions were fulfilled, we obtained $V_{A1,B1}=95.7(\pm 0.9)\%$, $V_{A2,B2}=94.2(\pm 0.6)\%$, $V_{A1,B2}=4(\pm 2)\%$ and $V_{A2,B1}=5(\pm 3)\%$ at a photon pair-rate \footnote{Sum of all detector combinations between Alice and Bob} of $21900(\pm 400)$/s by averaging the visibilities over $\SI{100}{s}$ while the alignment program was still running. Our achieved visibilities were limited by the state fidelity of our entangled photon source, rather than by our homemade alignment software and the polarization controllers. It is noteworthy that in general the visibility in the computational basis is higher than in the superposition basis. Our alignment procedure does not scan specifically for the computational basis as we do not control the polarization controller for the first measurement basis at Alice (PCA1), and therefore end up most likely in arbitrary superposition bases.

\begin{figure}[ht]
\begin{center}
\includegraphics[width=0.5\textwidth,clip]{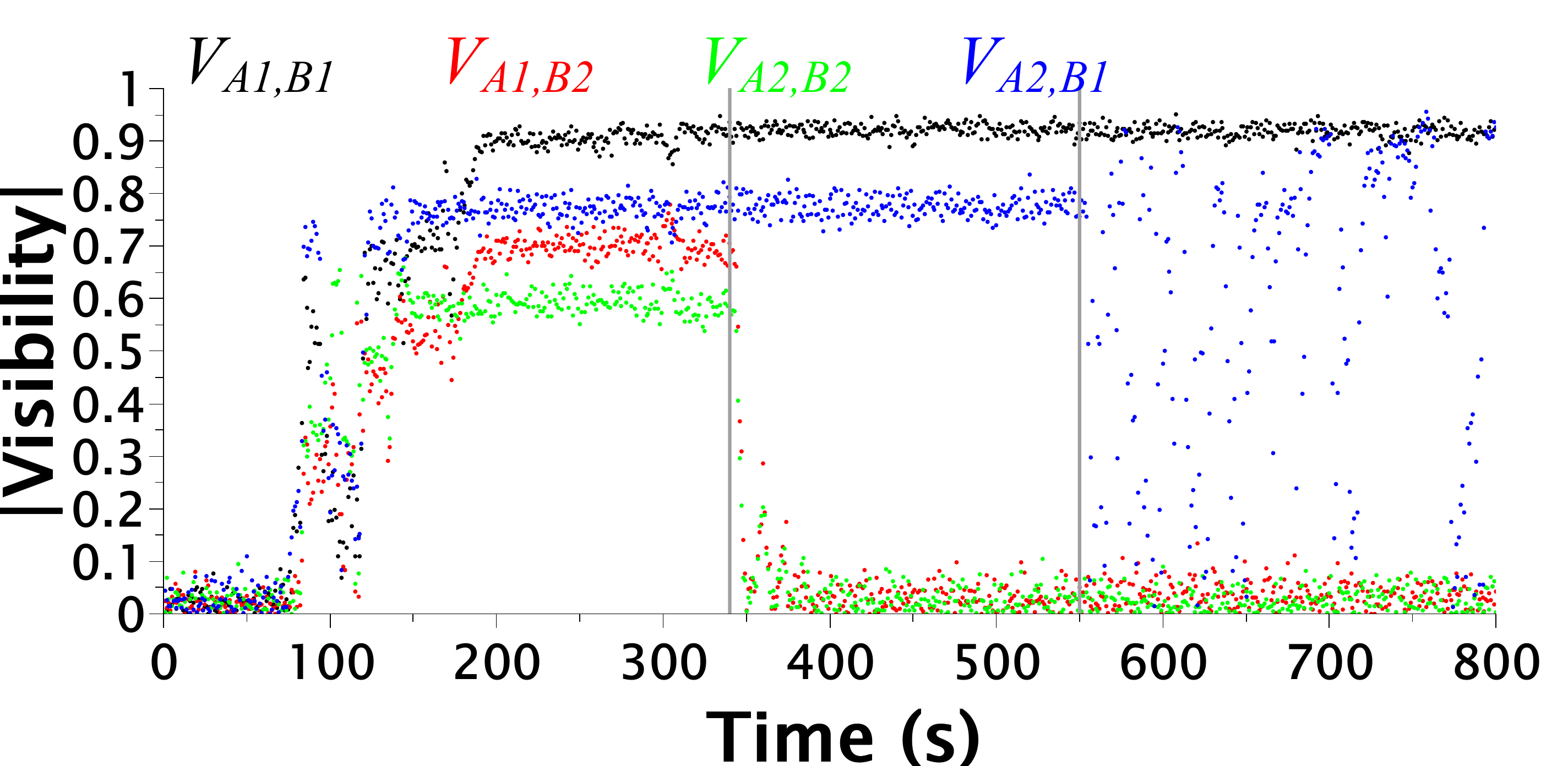}
\end{center}
\caption{A basic alignment procedure which fails as Alice and Bob do not share an entangled state. Step 1 and 2 can be completed, but step 3 (getting a high visibility $|V_{A2,B2}|$ in the superposition basis) always fails. Here we plot the absolute value of each visibility. The three steps are separated by vertical grey lines.
}
\label{absolut_fail}
\end{figure}

An unsuccessful alignment attempt is shown in Fig.~\ref{absolut_fail}. In this case, Alice and Bob are sharing separable states. 
The alignment steps are executed with manual fiber polarization controllers in chronological order. The fist two alignment steps can still be completed successfully by setting A1 and B1 to $H/V$, resulting in a maximized $V_{A1,B1}$ (here both polarization controllers PCA1 and PCB2 need to be iteratively adjusted). The second alignment step is completed by setting B2 to any basis lying in the $R-L-D-A$ plane. During the third step, A2 can be set to be parallel to B2, but no visibility between those bases is observed, as Alice and Bob do not share an entangled state. The only visible effect during this alignment step is the fluctuation of $V_{A2,B1}$ (blue), indicating how close A2 is to $H/V$. In case the shared state might be separable, the procedure will simply fail and the sum $V_{A1,B1}+V_{A2,B2}\leq1$ can be seen as an entanglement witness \cite{GUHNE20091}.

\begin{figure}[!ht]
\begin{center}
\includegraphics[width=0.5\textwidth,clip]{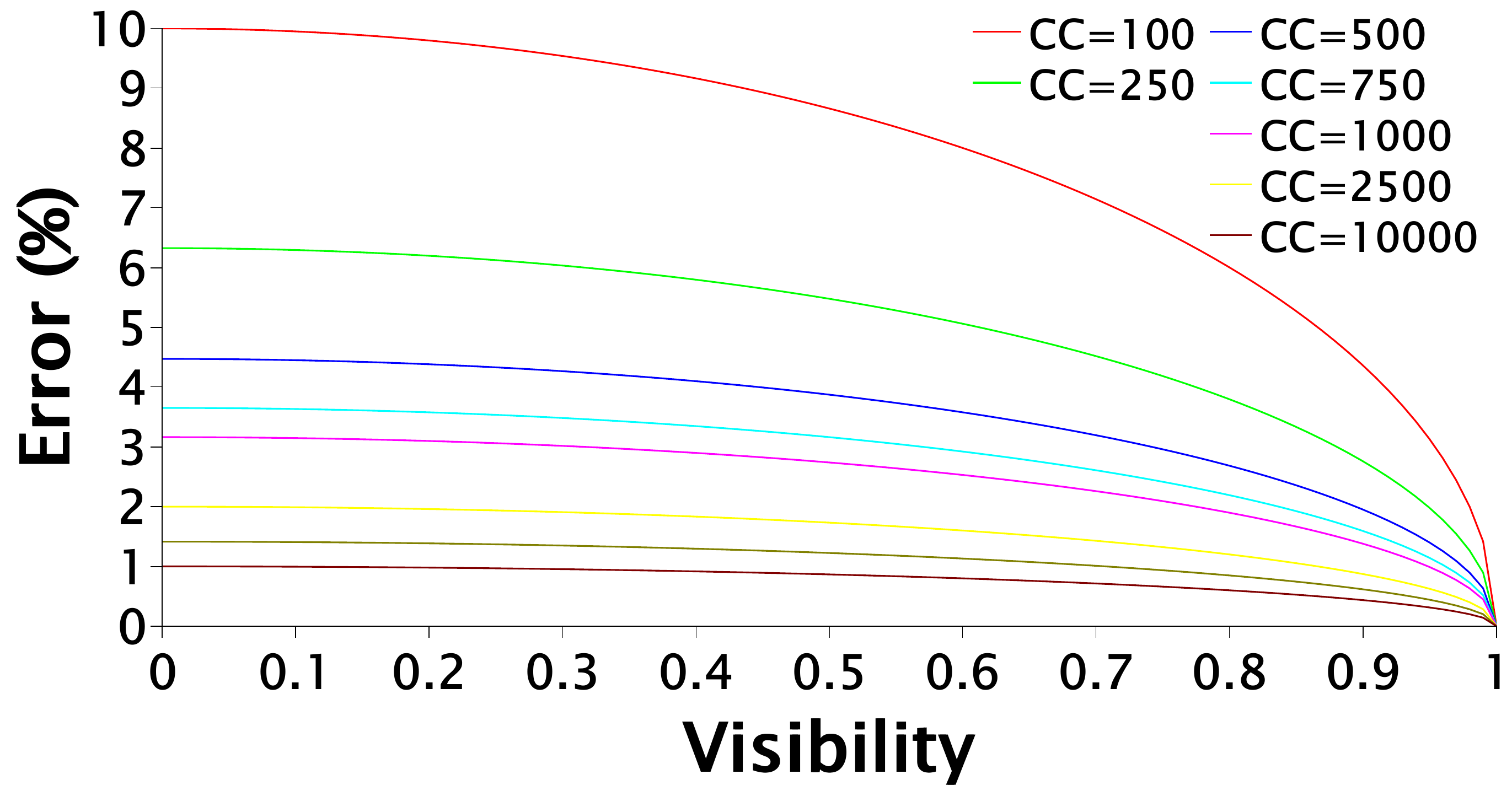}
\end{center}
\caption{Error propagation for the visibility formula in Eq. \ref{eqvis} assuming Poisson statistics for different total numbers of coincidence counts (CC).}
\label{error_calc}
\end{figure}

\section{Discussion}

We demonstrated an alignment and stabilization procedure for the required correlated and uncorrelated measurement bases for a QKD system solely based on the visibility (QBER). No information about the entangled state and the measurement bases is required. No additional polarizers nor alignment lasers are necessary, thus reducing the cost and complexity of QKD systems, therefore representing a step towards technological maturity. The alignment procedure is done on-the-fly, hence no time-consuming interruption of key distribution is required. However, a fraction $f$ of the distributed photon pairs is lost for key distillation because the measurement results have to be communicated to the other party to calculate the QBER in order to perform the alignment procedure. During the initial alignment phase as described in section \ref{sec:methods}, the fraction $f$ does not need to be minimized, as no secure key could be distilled at this point anyways and one can use all photon pairs available. Once these alignment conditions are fulfilled, a secure key can be generated and $f$ should be minimized to a level that only just allows to stabilize the setup. Thereby, one is only limited by the timescale of polarization fluctuations inside the glass fibers due to temperature gradients or other disturbances ~\cite{Wengerowsky2020} and the statistics or accuracy of the visibility measurement/calculation. Fig.~\ref{error_calc} shows the error propagation for the visibility calculation for different total numbers of coincidence counts. For high visibilities one needs only a small number of coincidence counts to get an accurate estimation of the visibility. This gives a rough estimate on how many coincidence counts are necessary for a wanted level of alignment precision.
The alignment procedure was demonstrated for the BBM92 protocol but works also for prepare-and-measure QKD schemes (BB84 \cite{Bennett:1984wv}) and copes with biased basis choices~\cite{Lo2004}. For the BB84 case, the alignment can be understood in terms of the Klyshko advanced wave picture \cite{1994JETP...78..259B}: Instead of sharing some measurement outcomes as in the entangled case, here the sender (Alice) needs to communicate information about which of her four states was sent.
In the first step, the receiver (Bob) sets one measurement basis to be maximally correlated and therefore aligned with Alice's first basis (up to unitary evolution given by the environment). In the second step, Bob sets his second basis to be uncorrelated with and therefore mutually unbiased to Alice's first basis. In the third and last step it is Bob who sends information about his choice of the basis to Alice. Alice then sets her second basis to be correlated with Bob's second basis.

As a distinct feature of our alignment method, we do not only align on (ideally) perfect correlations in bases A1, B1 and A2, B2, but also require no correlations between A1 and B2. Imperfections in the setup lead to small residual visibilities ($|V_{A1,B2}|$ and $|V_{A2,B1}| \neq 0$), hence non-perfect mutually unbiased measurement bases. This needs to be taken into account in post-processing as additional information potentially available to an eavesdropper needs to be removed.

The entire alignment scheme can be adapted to a scenario in which $U_{A1}/U_{A2}$ and $U_{B1}/U_{B2}$ are well known to Alice and Bob (for example by using bulk optics after the 50/50 beam splitter). In this case, both local bases for Alice (Bob) can be already pre-aligned to ensure mutual unbiasedness and the procedure simplifies a lot. Only one polarization controller is then required that allows to manipulate either $U_A$ or $U_B$ in order to align Alice's bases with those of Bob.

This  work was  supported  by  the  Austrian  Academy  of  Sciences (OEAW), the University of Vienna via the project QUESS and the Austrian Federal Ministry of Science, Research and  Economy (BMWFW).


\bibliography{Biblio_paper}

\appendix

\section{Theoretical background}
\label{appendix}

In this appendix we provide a theory support for the claims made in the main text, while we restrict our discussion to the case when the source emits pure two-qubit states $\ket{\psi}$. 
As illustrated in Fig.~\ref{fig1}, Alice and Bob then share the state:
\begin{align}
\ket{\psi_{i,j}} & = \left(U_{Ai}U_A \otimes U_{Bj}U_B\right)\ket{\psi} \\
& \equiv \left( \overline{U}_{Ai}\otimes \overline{U}_{Bj}\right)\ket{\psi},
\end{align}
which is subsequently subjected to local polarization measurements. These can be without loss of generality fixed to be measurements in the eigenbasis of Pauli $Z$ operator since operators $\overline{U}_{Ai}$ and $\overline{U}_{Bj}$ might be arbitrary. The visibility introduced in Eq.~\ref{eqvis} in the main text can then be identified with the expectation value of Pauli $Z$ measurement:
\begin{equation}
E^{\psi_{i,j}}=\bra{\psi_{i,j}}\sigma_z \otimes \sigma_z \ket{\psi_{i,j}}.
\end{equation}
In the following, we demonstrate the feasibility of the alignment procedure by showing that there indeed exist unitaries $U_{B1}$, $U_{B2}$ and $U_{A2}$ such that $E^{\psi_{1,1}} = E^{\psi_{2,2}} = 1$ and $E^{\psi_{1,2}} = 0$ and that these conditions automatically lead to $E^{\psi_{2,1}} = 0$. At last we show that by fulfilling all alignment conditions, Alice's two bases will always end up mutually unbiased to each other and the same is true for Bob's bases.

\subsection{Expectation values}

At first we draw the link between $E^{\psi_{i,j}}$ and the corresponding unitaries. Let the source produce a maximally entangled state $\ket{\psi}$. Then there exists a unitary $V$ such that
\begin{equation}
\label{sourcestate}
\ket{\psi}=\left( I\otimes V\right)\ket{\psi^{-}},
\end{equation}
where the singlet state $\ket{\psi^{-}}$ has the useful property that it remains invariant  when the same unitary transformation is applied to both sides:
\begin{equation}
\left(U\otimes U\right)\ket{\psi^{-}}=e^{i\omega}\ket{\psi^{-}}.
\end{equation}
This allows us to rewrite Alice's and Bob's state in the following way:
\begin{equation}
\begin{aligned}
\label{rewritten}
\ket{\psi_{i,j}} & = \left( \overline{U}_{Ai}\otimes \overline{U}_{Bj}V\right)\ket{\psi^{-}} \\
& = \left(I\otimes \overline{U}_{Bj}V\overline{U}_{Ai}^{\dag}\right)\left(\overline{U}_{Ai}\otimes \overline{U}_{Ai}\right)\ket{\psi^{-}} \\
& = e^{i\omega_i}\left(I\otimes U_\Delta^{i,j}\right)\ket{\psi^{-}},
\end{aligned}
\end{equation}
where we defined:
\begin{equation}
\label{uDelta}
U_\Delta^{i,j}=\overline{U}_{Bj}V\overline{U}_{Ai}^{\dag}.
\end{equation}
Using Eq.~\ref{rewritten} the expectation values can be rewritten as follows:
\begin{equation}
\label{expvalue}
\begin{aligned}
&E^{\psi_{i,j}}=\bra{\psi_{i,j}}\sigma_z \otimes \sigma_z \ket{\psi_{i,j}} \\
& = \bra{\psi^{-}}\left(I\otimes U_\Delta^{i,j\dag}\right)\left(\sigma_z \otimes \sigma_z\right)\left(I\otimes U_\Delta^{i,j}\right)\ket{\psi^{-}} \\
& = \bra{\psi^{-}}\left(I\otimes \left(U_\Delta^{i,j\dag}\sigma_zU_\Delta^{i,j}\sigma_z\right)\right)\left(\sigma_z\otimes \sigma_z\right)\ket{\psi^{-}} \\
& = -\bra{\psi^{-}}\left(I\otimes S_{ij}\right)\ket{\psi^{-}},
\end{aligned}
\end{equation}
where we defined:
\begin{equation}
S_{i,j}=U_\Delta^{i,j\dag}\sigma_zU_\Delta^{i,j}\sigma_z.
\end{equation}
We can use a general parametrization for an arbitrary unitary matrix $U$,
\begin{equation}
\label{unitary}
U=e^{i\alpha}\begin{pmatrix}
e^{-i\frac{(\beta+\delta)}{2}}\cos(\frac{\gamma}{2}) & -e^{-i\frac{(\beta-\delta)}{2}}\sin(\frac{\gamma}{2}) \\ e^{i\frac{(\beta-\delta)}{2}}\sin(\frac{\gamma}{2}) & e^{i\frac{(\beta+\delta)}{2}}\cos(\frac{\gamma}{2})
\end{pmatrix},
\end{equation}
to express $U_\Delta^{i,j}$ and calculate the explicit form of $S_{i,j}$:
\begin{equation}
S_{i,j}=\begin{pmatrix}
\cos(\gamma_{i,j}) & -e^{i\delta_{i,j}}\sin(\gamma_{i,j}) \\ -e^{-i\delta_{i,j}}\sin(\gamma_{i,j}) &\cos(\gamma_{i,j})
\end{pmatrix}.
\end{equation}
If we now plug this expression into Eq.~\ref{expvalue}, the expectation value takes a very simple form:
\begin{equation}
\label{psimexp}
E^{\psi_{i,j}}=-\cos\left(\gamma_{i,j}\right).
\end{equation}
This formula together with Eq.~\ref{uDelta} represents the link between measured visibilities and the form of measurement bases.

\subsection{Feasibility}

Considering the alignment procedure described in the main text, Bob can adjust PCB1 in such a way that $E^{\psi_{1,1}}=-1$ by setting $\gamma_{1,1}=0$, without knowing what this actually means in terms of his own unitary transformation $U_{B1}$, only based on observing the measured expectation value (visibility). Likewise, the conditions $E^{\psi_{1,2}}=0$ and $E^{\psi_{2,2}}=-1$ can be fulfilled by choosing $\gamma_{1,2}=\pi/2$ and $\gamma_{2,2}=0$ respectively. By inserting the three values $\gamma_{1,1}$, $\gamma_{2,2}$, and $\gamma_{1,2}$ back into $U_\Delta^{i,j}$ we get:
\begin{equation}
\label{eq:udelta}
\begin{aligned}
&U_\Delta^{i,i}=e^{i\alpha_{i,i}}\begin{pmatrix}
e^{-i\zeta_{i,i}} & 0 \\ 0 &e^{i\zeta_{i,i}}
\end{pmatrix} \qquad \text{and} \\&
U_\Delta^{1,2}=\frac{e^{i\alpha_{1,2}}}{\sqrt{2}}\begin{pmatrix}
e^{-i\zeta_{1,2}} & -e^{-i\eta_{1,2}} \\ e^{i\eta_{1,2}} &e^{i\zeta_{1,2}}
\end{pmatrix}
\end{aligned}
\end{equation}
with real parameters $\zeta_{i,j}=(\beta_{i,j}+\delta_{i,j})/2$ and $\eta_{i,j}=(\beta_{i,j}-\delta_{i,j})/2$. We can use Eq.~\ref{uDelta} to express Bob's unitaries $\overline{U}_{Bj}$ in terms of Alice's unitaries $\overline{U}_{Ai}$:
\begin{equation}
\label{Bjform}
\overline{U}_{Bj}=U_\Delta^{i,j}\overline{U}_{Ai}V^\dag.
\end{equation}
This way, Bob's unitaries are fully determined from \ref{eq:udelta} and $\overline{U}_{A1}$. The form of $\overline{U}_{A2}$ is determined in the next section.

\subsection{Vanishing cross-correlation}

Next we need to confirm that fulfilling conditions $E^{\psi_{1,1}} = E^{\psi_{2,2}} = -1$ and $E^{\psi_{1,2}} = 0$ forces $E^{\psi_{2,1}}=0$ as well. First, let us emphasize that Eq.~\ref{Bjform} are actually four different matrix equations. By reducing $\overline{U}_{Bj}$ from them one can simplify the rest into:
\begin{equation}
\label{ua2}
\overline{U}_{A2}=U_\Delta^{2,1\dag} U_\Delta^{1,1}\overline{U}_{A1}=U_\Delta^{2,2\dag} U_\Delta^{1,2}\overline{U}_{A1}.
\end{equation}
Solving this for the unknown matrix $U_\Delta^{2,1}$ yields:
\begin{equation}
U_\Delta^{2,1}=U_\Delta^{1,1}U_\Delta^{1,2\dag}U_\Delta^{2,2}=
\frac{e^{i\alpha_{21}}}{\sqrt{2}}\begin{pmatrix}
e^{-i\zeta_{21}} & -e^{-i\eta_{21}} \\ e^{i\eta_{21}} &e^{i\zeta_{21}}
\end{pmatrix}
\end{equation}
with substitutions $\alpha_{21}=\alpha_{11}-\alpha_{12}+\alpha_{22}$, $\zeta_{21}=\zeta_{11}-\zeta_{12}+\zeta_{22}$ and $\eta_{21}=\zeta_{11}-\zeta_{22}+\eta_{12}+\pi$. The structure of $U_\Delta^{2,1}$ is identical to that of $U_\Delta^{1,2}$ and so we can conclude that $\gamma_{2,1} = \pi/2$ and $E^{\psi_{2,1}}=0$. From Eq.~\ref{ua2} one also retrieves the form of unitary $\overline{U}_{A2}$.

\subsection{Mutual unbiasedness}

In order to ensure security of a QKD setup, we also prove that Alice and Bob measure in two mutually unbiased bases. We present the proof for Alice's bases, the calculation for Bob is analogous. We define $\overline{U}_{\Delta A}=U_{A2} U_{A1}^\dag$, which represents the transformation a photon would undergo when travelling from A1 to A2. Using Eq.~\ref{ua2} this can be expressed as follows:
\begin{equation}
\begin{aligned}
\overline{U}_{\Delta A} & = \overline{U}_{A2} \overline{U}_{A1}^\dag = U_{\Delta}^{2,2\dag}U_{\Delta}^{1,2} \\
& = \frac{e^{i(\alpha_{12}-\alpha_{22})}}{\sqrt{2}}\begin{pmatrix}
e^{-i(\zeta_{12}-\zeta_{22})} & -e^{-i(\eta_{12}-\zeta_{22})} \\ e^{i(\eta_{12}-\zeta_{22})} &e^{i(\zeta_{12}-\zeta_{22})}
\end{pmatrix}.
\end{aligned}
\end{equation}
Note that every component of this matrix has a modulus squared of $1/2$ and that Alice's measurement bases A1 and A2 are defined as $\{ \overline{U}_{A1}^\dag\ket{H} , \overline{U}_{A1}^\dag\ket{V}\}$ and $\{ \overline{U}_{A2}^\dag\ket{H} , \overline{U}_{A2}^\dag\ket{V}\}$, respectively. From there it follows that the overlap of any two vectors $\ket{\chi_l} \in \mathrm{A1}$ and $\ket{\phi_k} \in \mathrm{A2}$ reads:
\begin{equation}
|\braket{\phi_{k}}{\chi_{l}}|^2 = |\bra{k} \overline{U}_{\Delta A} \ket{l}|^2=\frac{1}{2}
\end{equation}
for all $k,l \in \{H,V\}$. Bases A1 and A2 are thus mutually unbiased.

Note that when investigating similar relations between Alice's and Bob's bases, Eq.~\ref{Bjform} leads to:
\begin{equation}
|\bra{k} \overline{U}_{Bj} \overline{U}_{Ai}^\dag\ket{l}|^2 = |\bra{k} U_\Delta^{i,j}\overline{U}_{Ai}V^\dag \overline{U}_{Ai}^\dag \ket{l}|^2.
\end{equation}
Only if $V=I$ does the overlap of both bases always reduce to expressions that depend only on $U_\Delta^{i,j}$:
\begin{equation}
|\bra{k_{Ai}}U_\Delta^{i,j}\ket{l_{Bj}}|^2 = 
\begin{cases}
1 & i = j \\
1/2 & i \neq j \\
\end{cases}
\end{equation}
for $i,j\in \{1,2\}$ and $k,l \in \{H,V\}$. This means that only in the case when the source emits a $\ket{\psi^-}$ state are Alice and Bob's bases guaranteed to be aligned with respect to each other or mutually unbiased (depending on the correlations). If any other maximally entangled state is used, this might no longer be the case. Two uncorrelated bases are then not necessarily mutually unbiased and even two identical measurement bases do not need to be correlated. However, the whole procedure still works and Alice's (Bob's) two bases will be mutually unbiased with respect to each other after the procedure is finished.

\end{document}